\documentstyle[prl,aps,twocolumn]{revtex}
\newcommand{\cnk}{C_N^k}
\newcommand{\cnm}{C_N^m}
\newcommand{\cnn}{C_N^n}
\newcommand{\cnl}{C_N^l}
\newcommand{\ro}{\rho(r,\varphi)}
\begin{document}
\title{Time-reversal symmetry and random polynomials}
\author{Daniel Braun$^1$, Marek Ku\'s$^{1,2}$ and Karol
\.Zyczkowski$^{1,3}$}
\address{$^1$Fachbereich Physik, Universit\"at-GH Essen, 45117
Essen, Germany \\
$^2$Center for Theoretical Physics, Polish Academy of Sciences,
Warsaw, Poland \\
$^3$Instytut Fizyki, Uniwersytet Jagiello\'nski, Krak\'ow, Poland}
\maketitle
\begin{abstract}
We analyze the density of roots of random polynomials where each
complex coefficient is constructed of a random modulus and a fixed,
deterministic phase.
The density of roots is shown to possess a singular component only in
the case
for which the phases increase linearly with the index of coefficients.
This means that, contrary to earlier belief, eigenvectors of a typical
quantum chaotic system with some antiunitary
symmetry will {\sl not} display a clustering
curve in the stellar representation. Moreover, a class  of
time-reverse invariant quantum systems is shown, for which spectra
display fluctuations characteristic of
orthogonal ensemble, while eigenvectors confer to
predictions of unitary ensemble.
\end{abstract}

The distribution of roots of polynomials of high degree with random
coefficients was investigated recently in connection with properties
of quantum chaotic systems \cite{leboeuf:90,bogomolny:92,bogomolny:96,LS96}.
In particular the authors of the cited references considered the
coherent state representation of eigenstates of a quantum mechanical
spin system with the total spin $S$. The polynomials in question
have the form
\begin{equation}
P(z)=\sum_{k=0}^{N}\sqrt{C_N^k} a_kz^k, \quad N=2S
\label{poly}
\end{equation}
where
 $C_N^k$ stand for binomial coefficients and
 $a_k$ are components of an eigenvector. The complex
variable $z$ is connected to the Bloch sphere angular variables
$\theta,\phi$ {\it via} $z=\tan(\theta/2)\exp(i\phi)$. It was shown by
Leb{\oe}uf and Voros \cite{leboeuf:90}, that for large values of $S$
when the quantum system in question is chaotic the distribution of
the roots is given by
\begin{equation}
\rho(z)=\frac{N}{\pi}\frac{1}{1+|z|^2}
\label{uni}
\end{equation}
corresponding to the uniform distribution of the roots over the Bloch
sphere. This is the consequence of the fact, that in the semiclassical
limit $N\to\infty$ the components with respect to a ``generic basis''
of the eigenvectors of a chaotic systems are independently normally
distributed (see \cite{haake:91} and references therein).

The details of the distribution of the components $a_k$ depend on
symmetries of the system in question. For systems which are not
time-reversal invariant the eigenvector components are complex,
with independently, normally distributed real and imaginary parts,
whereas for time-reversal invariant systems the eigenvectors can be
made real (also with normally distributed components).  In the
latter situation the uniform distribution (\ref{uni}) is modified.
In particular, the roots tend to concentrate on the real line
{\rm Im}$z=0$, which is a symmetry line for the roots (if $z_0$ is a
root then it's complex conjugate $z_0^*$ is also a root)
\cite{bogomolny:92,bogomolny:96}, see also below). When projected
back on the sphere the symmetry line is the great circle $\phi=0$.

This simplest situation corresponds to the case when  the
time-reversal operator is represented by the complex conjugation
operator.  On the other hand it is known that generalized
time-reversal symmetries, represented by the complex conjugation
supplemented by a unitary transformation, influence statistical
properties of eigenvector components in the same way as the
conventional time-reversal symmetry \cite{haake:91}. As an
illustration the authors of Refs.\ \cite{bogomolny:92,bogomolny:96}
considered various models of the so called kicked top system
\cite{kus:87}, which is described by the one-step evolution operator
of the form $U=\exp(-if_1)\exp(-if_2)\exp(-if_3)$ with
$f_i=f_i(S_x,S_y,S_z), i=1,2,3$ polynomial functions of the
components of the spin operator ${\bf S}=(S_x,S_y,S_z)$.  The
simplest case displaying chaotic dynamics in the classical limit is
obtained by choosing
$U_0=\exp(-i \mu S_x)\exp(-i p S_z^2/2S)$ with
appropriate values of the parameters $\mu$ and $p$.
It has two generalized time-reversal
symmetries $T_1=\exp(-i\mu S_x)K$ and $T_2=\exp(-i\mu S_x)\exp(i\pi
S_y)\exp(i\pi S_z)K$, $T_iUT_i^{-1}=U^\dagger$
both being compositions of linear rotations with the complex
conjugation operator $K$. The rotations shift the symmetry
line from the great circle $\phi=0$ to other ones, the phenomenon
exhibited by the numerical investigations performed by the authors
of Refs.\ \cite{bogomolny:92,bogomolny:96}.

A non-homogeneous distribution of zeros of Husimi functions is linked to
statistical properties of coherent states expanded in the eigenbasis of the
Floquet operator. In particular, the number of relevant eigenstates
\cite{HKS86} and the entropy of coherent states \cite{kz90} was found for
this model to be smaller than average along the symmetry lines $T_i$. A smaller
number of significantly occupied eigenstates denotes a larger number of weakly
occupied states, in consistency with investigated clustering of zeros of
eigenstates in Husimi representation along the symmetry curves.  Moreover,
the distribution of expansion coefficients of a coherent state localized
sufficiently far away from the symmetry lines is statistically
indistinguishable from properties of a generic coherent state of a system
without any antiunitary symmetry \cite{kz90}. This corresponds to recent
result of Prosen \cite{Pr96}, who showed that the densities of zeros of
random polynomials with real and complex coefficient are equal sufficiently
far away from the real axis.

In order to break the generalized time-reversal symmetry the
original model $U_0$ was supplemented by an nonlinear rotation $f_1=q
S_y^2/2S$ (in Ref.\
\cite{bogomolny:92}) or $f_1'=qS_z^2/2S$ (in Ref.\ \cite{bogomolny:96})
instead of $f_1=0$.  In their numerical investigations Bogomolny et
al.\ observed
vanishing of the concentration of the roots which they attributed to
the breaking of the time-reversal symmetry.  In what follows we will
argue that the concentration of the roots on the symmetry lines
happens in the case of generalized time invariance only
exceptionally and as such cannot be treated as a criterion
discriminating between the time reversal invariant and noninvariant
systems.  In particular the kicked tops
$U_1=\exp(-i q S_y^2)U_0$ and $U_2=\exp(-i q S_z^2)U_0$
differ with respect to the statistical properties of the spectra
for generic values of the parameter $q$.
The additional rotation term brakes all
generalized time-reversal
symmetries  for the first top and $U_1$ pertains to the circular unitary
ensemble (CUE),  while the second still possess such a symmetry
\begin{equation} \label{symetry}
T' =\exp(-iqS_z^2)\exp(-i\mu S_x)\exp(-iqS_z^2)K,
\end{equation}
 and its spectrum is typical to circular orthogonal
ensemble (COE), irrespective of the value of $q$.
Note that the above operator is constructed of a nonlinear unitary
rotation (quadratic term $S_z^2$ in the exponent), in contrast to the
operators $T_1$ and $T_2$.

Inasmuch as level statistics reveals directly the symmetry
properties of quantum systems, special care has to be taken
interpreting the statistical properties of eigenvectors,
since their distribution depends on the basis chosen.
For example, the distribution of eigenvectors of $U_0$ in the
$S_z$ basis does not confer to COE predictions. The agreement
with random matrices is recovered in $S_x$ basis: the geometric
symmetry
of the top manifests itself in the structure of operator $U_0$. It
splits into two parities of size $S$ and $S+1$, which have to be
treated separately to achieve results according to random
matrices. In earlier references \cite{KMH,HZ,LZ} the variables of the
top were exchanged
$x \leftrightarrow z$, what gives the same effect.

The distribution of eigenvectors can be characterized by their mean
entropy $H$, which for random matrices of size
$N$ is equal to
$H(N,\beta)=\Psi(N\beta/2+1)-\Psi(\beta/2+1)$, where $\Psi$ stands
for the digamma function and  $\beta=1$ for COE and  $\beta=2$ for CUE
       \cite{jones}.
Figure 1 presents the entropy of eigenvectors relative to the entropy of
CUE for two tops $U_1$ and $U_2$ as a function of the control parameter
$q$. Observe similar behaviour for "unitary" top $U_1$ and the
"orthogonal" top $U_2$! The dips in the data for unitary top at $q=0$
and $q=p=6.0$ correspond to transitions to the orthogonal class, while
 $U_2$ pertains to COE for any value of $q$
due to the symmetry (\ref{symetry}). This difference is visualized in
level spacing distribution $P(s)$ displayed in the inset.
An explanation of this fact is simple: out of any "orthogonal" spectrum
$D_1$ by a generic unitary rotation $W$ one can produce an
operator $U_W=WD_1W^{\dagger}$ which enjoys COE-like properties of the
spectrum and CUE-like properties of the eigenvectors. This is exactly
the case of the top $U_2$, for which the operator $\exp(-i q S_z^2)$
plays the role of $W$. Observe that $U_2$ is similar to the orthogonal
 top
$U_2'=\exp(-i \mu S_x)\exp[-i (p+q) S_z^2/2S]$.

A similar effect is visible in the distribution of zeros of Husimi
function representing eigenvectors: both tops show
lack of roots concentration lines as shown for $U_1$ in
\cite{bogomolny:92}
and for $U_2$ in  \cite{bogomolny:96}, even though they belong to
different universality classes.

In order to understand the above announced results let us
derive the density of roots $\rho$ of a polynomial (\ref{poly}),
where $a_{k}$ are  Gaussian distributed random quantities {\em with
fixed} but arbitrary phases $\varphi_k$:
\begin{equation} \label{eq:ak}
a_k=r_ke^{i\varphi_k}\,,
\end{equation}
the $r_k$ being distributed according to

\begin{equation} \label{eq:Prk}
P(r_k)=\frac{1}{\sqrt{2\pi}}e^{-r_k^2/2}\,.
\end{equation}

We will make use of the same technique employed in
Ref.\cite{bogomolny:96}, namely representing $\ro$ by the
Kac--formula,

\begin{equation} \label{eq:Kac}
\rho(z)=\delta [P(z)]\left|\frac{dP(z)}{dz}\right|^2\, ,
\end{equation}
 and then expressing the delta--functions for the real -- and
imaginary parts
of $P(z)$ as Fourier integrals. We then get, in full analogy with
Eq.(C6) in Ref.\cite{bogomolny:96}:

\begin{eqnarray}
\ro&=&\frac{1}{(2\pi)^2}\int d\xi_1\int
d\xi_2\Big\{\sum_{k=0}^Nk^2\cnk r_k^2r^{2(k-1)}+\nonumber\\
&&\sum_{k\ne
l=0}^Nkl\sqrt{\cnk\cnl}r_kr_l\,r^{k+l-2}e^{i(\varphi_k-\varphi_l)}
e^{i(k-l)\varphi}\Big\}\cdot   \nonumber\\ 
&&\cdot\exp{\sum_{k=0}^Nr_n(\alpha_n\cos\varphi_n+\beta_n\sin\varphi_n)}\,
\end{eqnarray}
where
\begin{eqnarray}
\alpha_n&=&ir^n\sqrt{\cnn}(\cos(n\varphi)\xi_1+\sin(n\varphi)\xi_2)\\
\beta_n &=&ir^n\sqrt{\cnn}(\cos(n\varphi)\xi_2-\sin(n\varphi)\xi_1)\,
\end{eqnarray}
and $z=re^{i\varphi}$.
Averaging over the random coefficients $r_k$ amounts now to simple
Gaussian
integrations. The resulting average density can be cast in the following
form:

\begin{eqnarray}
\langle\ro\rangle&=&\frac{1}{(2\pi)^2}\int d\xi_1\int d\xi_2
(A+B\xi_1\xi_2+C\xi_1^2+D\xi_2^2)\nonumber\\
&&\exp(-a\xi_1^2-b\xi_2^2-2c\xi_1\xi_2)\label{eq:xiints}\,,
\end{eqnarray}
where
\begin{eqnarray}
A&=&\sum_{k=0}^Nk^2r^{2(k-1)}\cnk\\
B&=&-\sum_{k,l}^Nh_{kl}\sin(\varphi_k+k\varphi+\varphi_l+l\varphi)\\
C&=&-\sum_{k,l}^Nh_{kl}\cos(\varphi_k+k\varphi)\cos(\varphi_l+l\varphi)\\
D&=&-\sum_{k,l}^Nh_{kl}\sin(\varphi_k+k\varphi)\sin(\varphi_l+l\varphi)\,,
\end{eqnarray}
and $h_{kl}=klr^{2(k+l-1)}\sqrt{\cnk\cnl}\cos(\varphi_k+k\varphi-\varphi_l
-l\varphi)$. The coefficients of the quadratic form in the exponential are
given by
\begin{eqnarray}
a&=&\frac{1}{2}\sum_{n=0}^N\cnn r^{2n}\cos^2(\varphi_n+n\varphi)\\
b&=&\frac{1}{2}\sum_{n=0}^N\cnn r^{2n}\sin^2(\varphi_n+n\varphi)\\
c&=&\frac{1}{4}\sum_{n=0}^N\cnn r^{2n}\sin(2(\varphi_n+n\varphi))\,.
\end{eqnarray}
It has the eigenvalues
\begin{equation}
\lambda_{1,2}=(a+b\pm\sqrt{(a-b)^2+4c^2})/2\,,
\end{equation}
and can be diagonalized by a rotation in the $(\xi_1,\xi_2)$ plane by an
angle $\gamma=-\arctan((a-b-\sqrt{(a-b)^2+4c^2})/2c)$. The
$\xi$--integrals
are Gaussian again and lead to the following  explicit expression for
the mean density of roots:

\begin{eqnarray}
\langle\ro\rangle&=&\frac{1}{4\pi}\bigg[\frac{A}{\sqrt{\lambda_1\lambda_2}}\nonumber\\
&&+\frac{1}{2}\left(\frac{K_1}{\sqrt{\lambda_1^3\lambda_2}}+\frac{K_2}{\sqrt{\lambda_1\lambda_2^3}}\right)+\nonumber\\
&&\frac{1}{4\pi}\frac{K_3}{\lambda_1\lambda_2}\bigg]\label{eq:rofinal}\,,
\end{eqnarray}
with the coefficients

\begin{eqnarray}
K_1&=&-B\cos\gamma\sin\gamma+C\cos^2\gamma+D\sin^2\gamma\\
K_2&=&+B\cos\gamma\sin\gamma+C\sin^2\gamma+D\cos^2\gamma\\
K_3&=&+B\cos 2\gamma+(C-D)\sin 2\gamma\,.
\end{eqnarray}
Obviously, $\langle\ro\rangle$ can only be singular if at least one
of the
two eigenvalues $\lambda_1$ or $\lambda_2$ is zero. This condition
leads to
$ab=c^2$. After some straightforward manipulations it can be
written in the
form $Q(r,\varphi)=0$ with

\begin{equation} \label{eq:cond}
Q(r,\varphi)=\sum_{n<m=0}^N
\cnn\cnm\sin^2(\varphi_m+m\varphi-\varphi_n-n\varphi)r^{2(n+m)}\,.
\end{equation}
Thus, the points $(r,\varphi)$ for which the average density of roots
diverges are the zeros of the polynomial in Eq.(\ref{eq:cond}). However,
$Q(r,\varphi)$ is positive semi--definite. The only possibility of
$Q(r\varphi)=0$ is given by $r=0$ (which is always a solution and thus
always a point of singular density), or by simultaneous vanishing of all
coefficients:
$\sin(\varphi_m+m\varphi-\varphi_n-n\varphi)=0$ for all $m,n$. In
the latter
case $Q(r,\varphi)$ will be zero for all $r$, implying immediately that
lines of singular density can be only straight lines in the
$z$--plane. On
the other hand, assuming that  $\varphi_k$ is a  differentiable function
of the index $k$, one finds that $\varphi_k=-k\varphi+const.$ with a
$k$--independent constant. Since the phases $\varphi_k$ were chosen as
constants,
the only way to fulfill this equation is by $\varphi_k=k\alpha+\beta$,
$\varphi=-\alpha$.
 For any other choice of the $k$--dependence of
the $\varphi_k$, lines with more or less pronounced maxima of $\ro$ may
still exist, but the singular character of the density is lost ---
with the exception of the origin.

 Above reasoning proves our claim that curves of singular density
are only possible if the phases $\varphi_k$ increase linearly with the
index $k$.  This is
exactly the case of the top $U_0$, for which the symmetries $T_1$ and
$T_2$ manifest themselves as singularities along straight lines on the
complex plane, which correspond to great circles on the sphere.

On the other hand, all the
deviations from the above form  result in a blurring of the sharp
lines seen
when plotting numerically obtained roots of random polynomials,
irrespective of whether a particular symmetry of the possibly
underlying physical system is still preserved or not.
To demonstrate this effect we have analyzed random polynomials
(\ref{poly}) with coefficients (\ref{eq:ak}) given by
$\varphi_k=qk^2/N$. This assumption  corresponds to the problem
induced by the
generalized time-reversal symmetry (\ref{symetry}) of the top $U_2$. For
$q=0$ (real coefficients $a_k$) the distribution of zeros suffers
a singularity along the real axis, while for
larger value of $q$ the clustering curve
 twists and acquires a finite width. This is visible in
figures 2 and 3 where we plotted on a complex plane zeros of 50 random
polynomials with $N=40$ (part
a) and the density of zeros obtained  according to Eq.(\ref{eq:rofinal})
(part b). For q=0.2 the symmetry line already deviates from the real axis.
For $q=0.5$ a ridge in the density of zeros is still
observed, at $q\sim 1$ the distribution of zeros is almost
homogeneous.
Interestingly, the qualitative character of the density does not change
much with $N$.

Let us mention here that the density of zeros of random polynomials
(\ref{poly}) with fixed phases can be obtained using  slightly
different techniques proposed by Edelman and Kostlan \cite{EK95},
 Shepp and Vanderbei \cite{shepp:95} or
Prosen \cite{Pr96}. Moreover, the density of roots of some generalized
random polynomials was recently discussed in \cite{MBFMA96}.

It is a pleasure to thank P. Leb{\oe}uf and T. Prosen for stimulating
discussions and for providing us their preprints prior to publication.
M.K. and K.\.Z. are grateful to F. Haake for hospitality during their
stays in Essen, where this work has been initiated. This work
was partially supported by the
Sonderforschungsbereich ``Unordnung und gro{\ss}e Fluktuationen'' der
Deutschen Forschungsgemeinschaft and partially
 by Polish Committee of
Scientific Research under the Grant No.~2P03B~03810.

\begin{figure}
\caption{Mean entropy of eigenvectors compared with the entropy $H_{CUE}$ of
the unitary ensemble  drawn as a function of the perturbation parameter
$q$ for two models: "unitary" top $U_1(\triangle)$ and "orthogonal" top
$U_2(\circ)$ with $\mu=1.7, p=6.0$ and spin length $S=40$.
The dashed line represents the value $H_{COE}/H_{CUE}\approx 0.91$. The inset
shows the cumulative level spacing distribution $P(s)$ obtained for both models
out of $100$ operators $U$ with fixed $q=2.0$ and $p$ varying from $6.0$ to
$12.0$ and compared to the Wigner surmises for both universality classes.
}
\end{figure}

\begin{figure}
\caption{The distribution of  roots of $50$ random polynomials with
quadratically 
increasing phases $(q=0.2)$ shown in part a)  follows the analytically
obtained density  shown in the contour plot in part b). The
concentration line of the 
zeros deviates from the real axis and is no longer a line of singular density.
}
\end{figure}

\begin{figure}
\caption{As in Fig 2 for $q=0.5$. The concentration line of the zeros
is even more blurred than for $q=0.2$.
}
\end{figure}
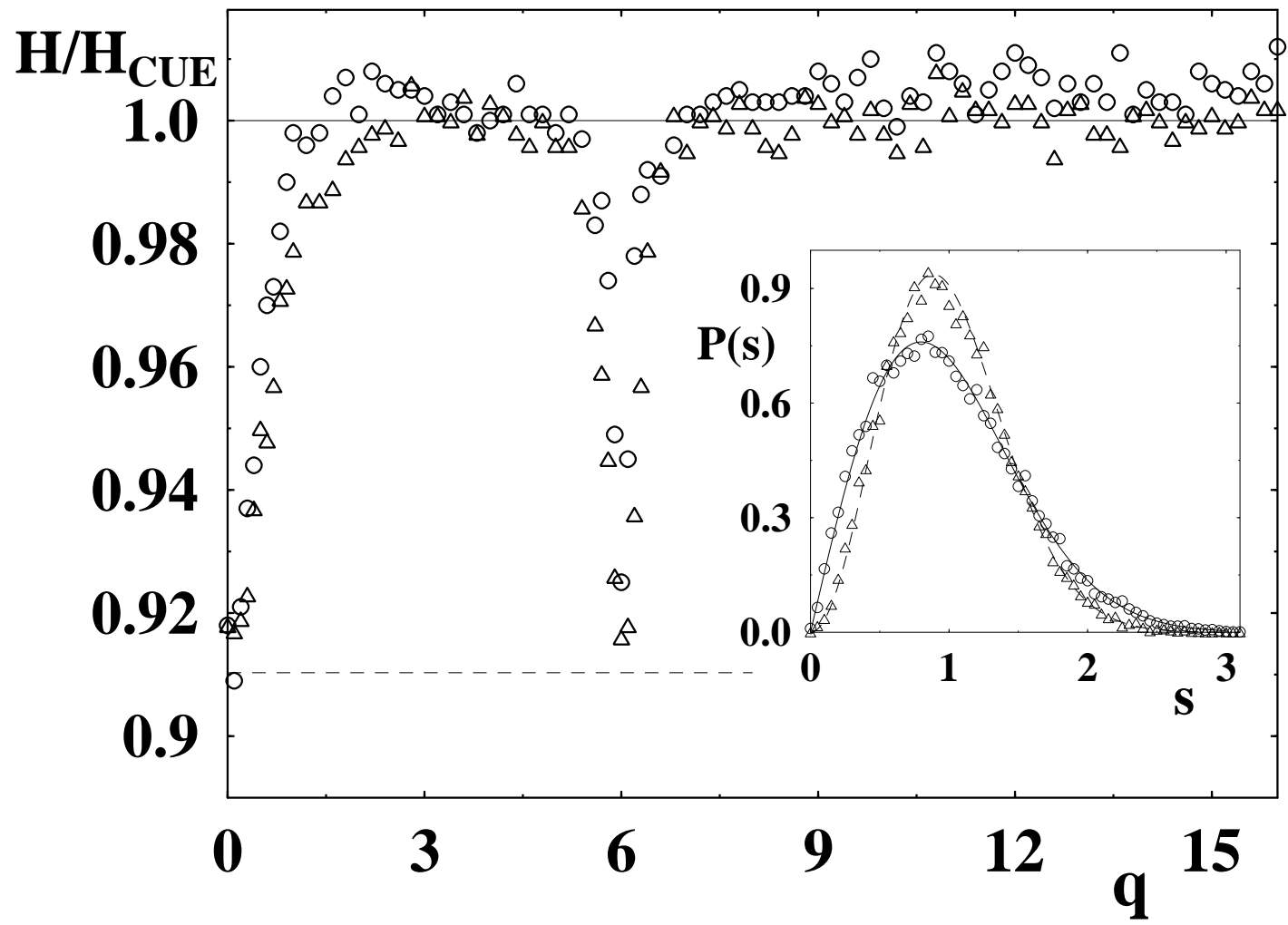
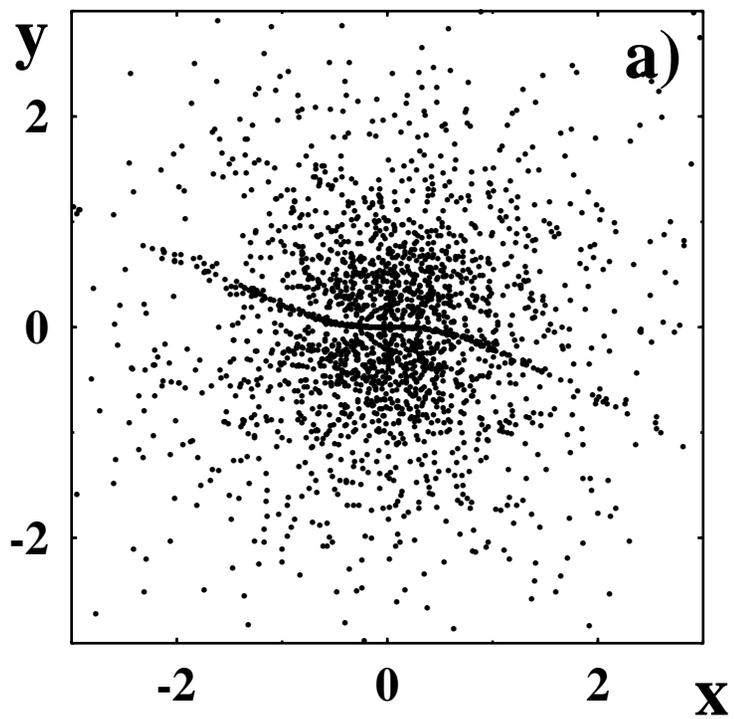
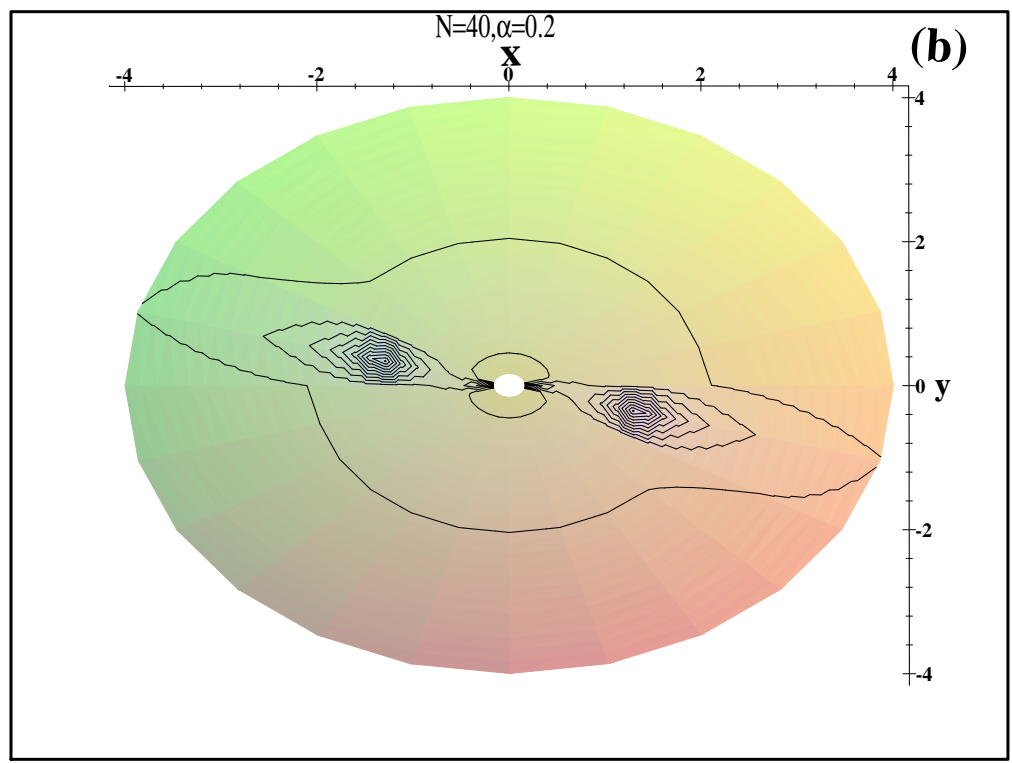
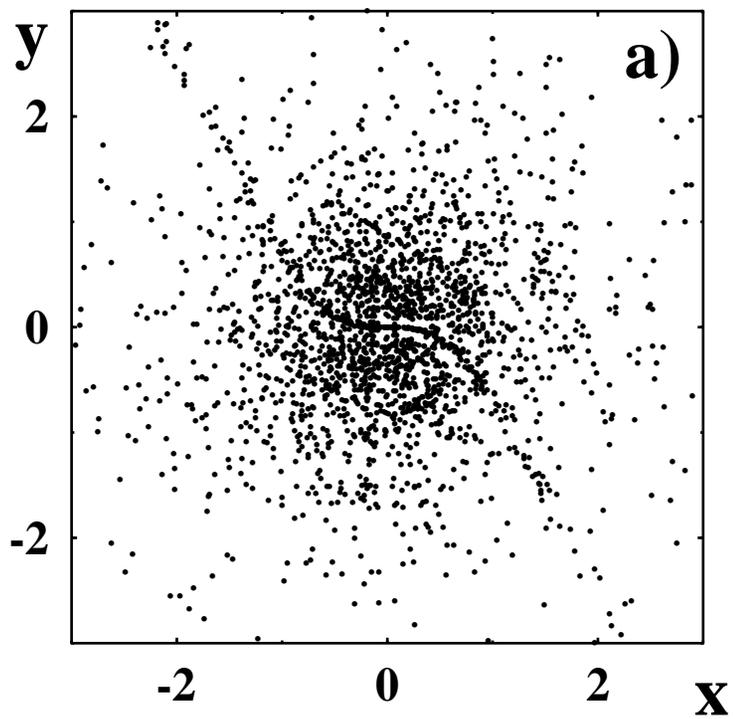
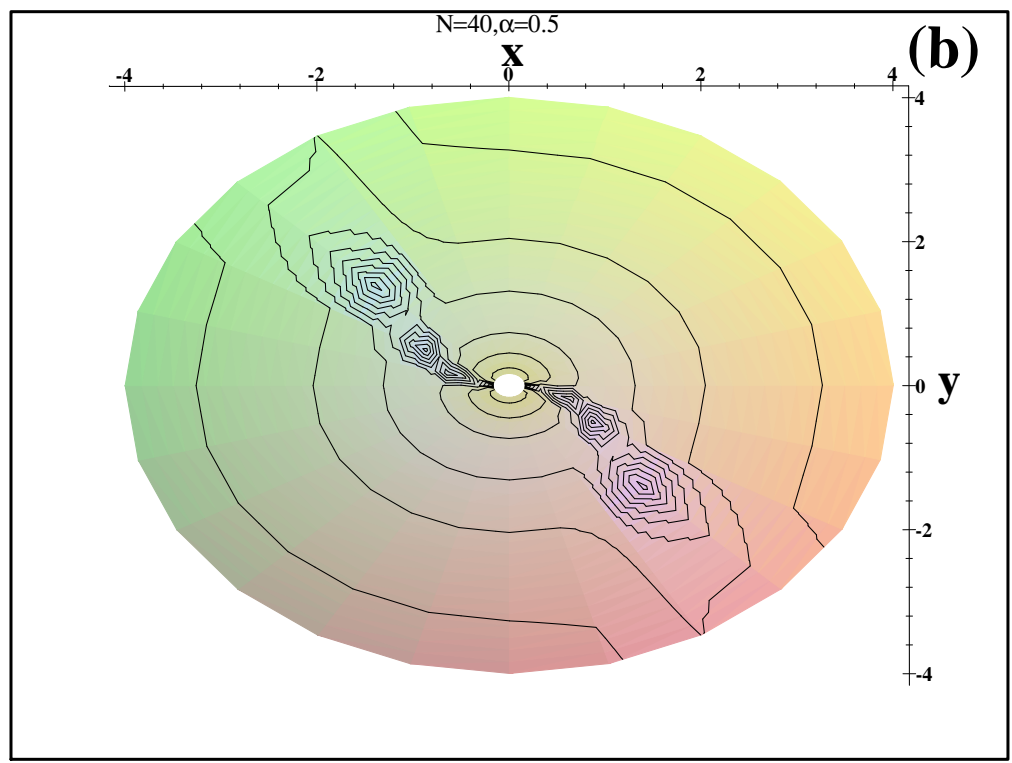
\end{document}